\documentclass[twocolumn,aps,epsf,superscriptaddress]{revtex4}
\usepackage[utf8]{inputenc}
\usepackage{amstext}
\usepackage{graphicx}

\makeatletter
\@ifundefined{textcolor}{}
{%
 \definecolor{BLACK}{gray}{0}
 \definecolor{WHITE}{gray}{1}
 \definecolor{RED}{rgb}{1,0,0}
 \definecolor{GREEN}{rgb}{0,1,0}
 \definecolor{BLUE}{rgb}{0,0,1}
 \definecolor{CYAN}{cmyk}{1,0,0,0}
 \definecolor{MAGENTA}{cmyk}{0,1,0,0}
 \definecolor{YELLOW}{cmyk}{0,0,1,0}
}

\@ifundefined{definecolor}{\usepackage{color}}{}
\@ifundefined{definecolor}{\usepackage{color}}{}
\@ifundefined{definecolor}{\usepackage{color}}{}

\usepackage{dcolumn}
\usepackage{bm}
\@ifundefined{definecolor}{\usepackage{color}}{} 
\usepackage{xcolor}

\makeatother

\begin{document}

\title{Unusual exchange couplings and intermediate temperature Weyl state in Co$_{3}$Sn$_{2}$S$_{2}$ }

\author{Qiang Zhang}
\email{zhangq6@ornl.gov}

\affiliation{Neutron Scattering Division, Oak Ridge National Laboratory, Oak Ridge, Tennessee 37831, USA}

\author{Satoshi Okamoto}
\email{okapon@ornl.gov}
\affiliation{Materials Science and Technology Division, Oak Ridge National Laboratory, Oak Ridge, Tennessee 37831, USA}
\affiliation{Quantum Science Center, Oak Ridge, Tennessee 37831, USA}

\author{German D. Samolyuk}
\affiliation{Materials Science and Technology Division, Oak Ridge National Laboratory, Oak Ridge, Tennessee 37831, USA}

\author{Matthew B. Stone}
\affiliation{Neutron Scattering Division, Oak Ridge National Laboratory, Oak Ridge, Tennessee 37831, USA}

\author{Alexander I. Kolesnikov}
\affiliation{Neutron Scattering Division, Oak Ridge National Laboratory, Oak Ridge, Tennessee 37831, USA}

\author{Rui Xue}
\affiliation{Department of Physics $\&$ Astronomy, The University of Tennessee, Knoxville, TN 37996, USA}

\author{Jiaqiang Yan}
\affiliation{Materials Science and Technology Division, Oak Ridge National Laboratory, Oak Ridge, Tennessee 37831, USA}

\author{Michael A. McGuire}
\affiliation{Materials Science and Technology Division, Oak Ridge National Laboratory, Oak Ridge, Tennessee 37831, USA}
\affiliation{Quantum Science Center, Oak Ridge, Tennessee 37831, USA}

\author{David Mandrus}

\affiliation{Department of Materials Science and Engineering, The University of Tennessee, Knoxville, TN 37996, USA}
\affiliation{Materials Science and Technology Division, Oak Ridge National Laboratory, Oak Ridge, Tennessee 37831, USA}
\affiliation{Department of Physics $\&$ Astronomy, The University of Tennessee, Knoxville, TN 37996, USA}

\author{D. Alan Tennant}
 
\affiliation{Materials Science and Technology Division, Oak Ridge National Laboratory, Oak Ridge, Tennessee 37831, USA}
\affiliation{Shull Wollan Center, Oak Ridge National Laboratory, Oak Ridge, Tennessee 37831, USA}
\affiliation{Quantum Science Center, Oak Ridge, Tennessee 37831, USA}

\date{\today}
\begin{abstract}

 Understanding the magnetism and its possible correlations to the topological properties has emerged as a forefront and difficult topic in studying magnetic Weyl semimetals. Co$_{3}$Sn$_{2}$S$_{2}$ is a newly discovered magnetic Weyl semimetal with a kagome lattice of cobalt ions and has triggered intense interest for rich fantastic phenomena. Here, we report the magnetic exchange couplings of Co$_{3}$Sn$_{2}$S$_{2}$ using inelastic neutron scattering and two density functional theory (DFT) based methods: constrained magnetism and multiple-scattering Green's function methods. Co$_{3}$Sn$_{2}$S$_{2}$ exhibits highly anisotropic magnon dispersions and linewidths below $T_{C}$, and paramagnetic excitations above $T_{C}$. The spin-wave spectra in the ferromagnetic ground state is well described by the dominant third-neighbor ``across-hexagon'' $J_{d}$ model. Our density functional theory calculations reveal that both the symmetry-allowed 120$^\circ$ antiferromagnetic orders support Weyl points in the intermediate temperature region, with distinct numbers and the locations of Weyl points. Our study highlights the important role Co$_{3}$Sn$_{2}$S$_{2}$ can play in advancing our understanding of kagome physics and exploring the interplay between magnetism and band topology.

\end{abstract}


 \maketitle
 The kagome lattice has attracted considerable interest since it can host new exotic magnetic and electronic states such as spin liquids\cite{Savary2017,Balents2012} and topological Dirac/Weyl fermions\cite{Ye2018,Balents2012,Liu2018}. Determination of the exchange couplings that control the magnetic phases is key to understanding these exotic states. While initial understanding of kagome magnetism is based on the isotropic nearest-neighbor (NN) exchange coupling\cite{Hastings2000,Yan2011}, recent research reveals a necessary requirement of considering the
 further-neighbor interactions\cite{He2015,Janson2008,Messio2011}. The kagome lattice with interactions beyond NN couplings tends to result in a very rich magnetic phase diagrams based on theoretical predictions \cite{Boldrin2015,Boyko2020,Messio2011,Bishop2010}. However, the experimental realizations of such  materials are rare. Two interesting examples are haydeeite MgCu$_{3}$(OH)$_{6}$Cl$_{2}$\cite{Janson2008,Boldrin2015,Janson2009} and Kapellasite ZnCu$_{3}$(OH)$_{6}$Cl$_{2}$\cite{Janson2008,Helton2010,Janson2009,Fak2012}, where the third-neighbor interaction competes with the NN one. Dominant third-neighbor exchange ``$J_{3}$'' was discovered in antiferromagnetic (AFM) BaCu$_{3}$V$_{2}$O$_{8}$(OD)$_{2}$ \cite{Boldrin2018} along with a very weak third-neighbor ``across-hexagon" $J_{d}$. Identifying new magnetic kagome materials with significant further-neighbor interaction is therefore of great interest to advance the field of kagome physics and to explore the emergent phenomena in such materials.

\begin{figure}
\centering \includegraphics[width=1\linewidth]{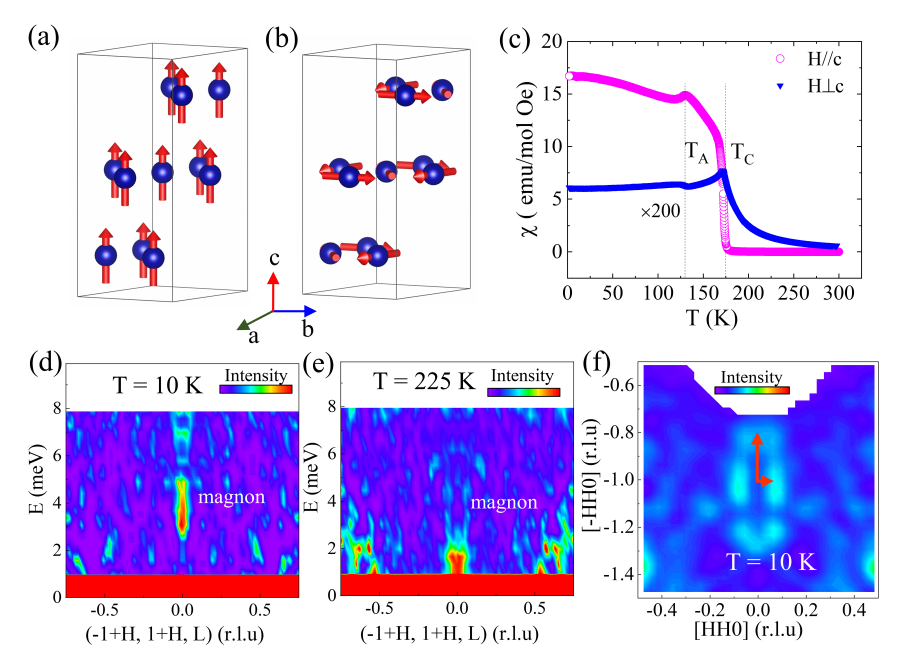} \caption{ (color online) Magnetic structures of Co$_{3}$Sn$_{2}$S$_{2}$: (a). FM order in $T<T_{A}$; (b). the proposed 120$^\circ$ AFM order in $T_{A}<T<T_{C}$ in Ref. \cite{Guguchia2020}; (c). The temperature dependence of the susceptibility with field parallel to $ab$ plane and $c$ axis in a field of 250 Oe.  Low energy magnetic excitations with E$_{i}$=15 meV near FM zone center ($-$1,1,1) at (d) 10 K and (e) 225 K. (f). Constant-E slice in the (H,H,0)-(-H,H,0) plane around E of 16 meV. }

\label{fig:structure} 
\end{figure}

The discovery of new magnetic Weyl semimetal Co$_{3}$Sn$_{2}$S$_{2}$ with a kagome lattice of cobalt ions has triggered tremendous interest to explore fantastic phenomena and underlying physics.\cite{Liu2018,Yin2019,Liu2019,Morali2019,Xu2018} Co$_{3}$Sn$_{2}$S$_{2}$ exhibits tunable magnetic states associated with anomalous Hall conductivity, and an exotic Weyl state at low temperatures \cite{Liu2018,Guguchia2020}. The ground state of  ferromagnetic (FM) order with moments along the $c$ axis exists below $T_{A}\approx130$ K, as shown in Fig.~\ref{fig:structure} (a). The FM order is believed \cite{Liu2018,Guguchia2020} to be important in the breaking of time reversal symmetry ($\mathcal{T}$) that induces a Weyl state in $T<T_{A}$ in Co$_{3}$Sn$_{2}$S$_{2}$. It is therefore crucial to determine the exchange couplings of the FM ground state. There have been theoretical attempts to determine these exchange couplings. It was proposed \cite{Legendre2020} that out-of-plane NN $J_{z}$ is mainly responsible for the formation of the FM order below $T_{A}$. In sharp contrast, Liu et al. \cite{Liu2020} employed a multiple-scattering Green's function method to calculate the exchange couplings and obtained a dominant NN in-plane $J_{1}$ to explain the FM order. These discrepancies show the difficulty in determining the precise exchange couplings and point to the need for careful experimental measurements to help constrain theoretical models and further our understanding and development of this material. Furthermore, recent investigations \cite{Kassem2017} revealed the existence of an anomalous phase showing peculiar properties like magnetic relaxation in the intermediate temperature region $T_{A}<T<T_{C}$. A coexistence of the ferromagnetic (FM) order and an AFM order was then uncovered by the $\mu$SR technique \cite{Guguchia2020}, with an in-plane 120$^\circ$ order proposed (see Fig.~\ref{fig:structure} (b)).  It would be very important to explore if there exists a correlation between different magnetic orders and band topology and whether Co$_{3}$Sn$_{2}$S$_{2}$ could host a new Weyl state in $T_{A}<T<T_{C}$.

In this letter, by a combined use of inelastic neutron scattering and linear spin wave theory, we find that the FM ground state of Co$_{3}$Sn$_{2}$S$_{2}$ is stabilized by the third-neighbor ``cross-hexagon" $J_{d}$, which makes Co$_{3}$Sn$_{2}$S$_{2}$ the first example of a kagome material exhibiting such an unusual magnetic interaction. This is compared with the exchange couplings extracted from our two density functional theory (DFT) based methods: one is a constrained magnetism method to compare the total energy of magnetically ordered states and the other utilizes multiple-scattering Green's function methods considering small spin reorientations from the magnetic ground state. Using DFT, we compare the electronic band structures associated with FM order and the two possible symmetry-allowed 120$^\circ$ AFM orders. Distinct Weyl points are found to exist for $T_{A}<T<T_{C}$ associated with the two different 120$^\circ$ AFM orders, indicating that a new Weyl state may exist in this temperature region.

The details of sample preparation and characterization, neutron scattering experiments, spin wave analysis using SpinW package, and DFT calculations can be found in the supplemental material (SM), which includes $Refs$.  \cite{May2020,Toth2015,Zhang2019,Kressee1999,Perdew1996,Kressee1996,Ebert2011,Ebert2017,Liechtenstein1995,wanniertools,wannier90}. Our neutron diffraction experiment at 10 K confirms the reported rhombohedral structure with space group of $\rm R\mbox{-}3m$ (No. 166) \cite{Vaqueiro2009,Liu2018,Schnelle2013,Kassem2017}. The temperature dependence of susceptibility of the single crystal in Fig.~\ref{fig:structure} (c) exhibits two anomalies at $T_{C}\approx$ 175 K and $T_{A}\approx$ 130 K. The field dependence of magnetization in Fig. S2 (b) shows an easy direction along the $c$ axis with a moment $\approx$ 0.37 $\mu_B$, consistent with the FM order reported previously \cite{Guguchia2020,Kassem2017}. In Fig.~\ref{fig:structure} (d-e), we compare the low-E magnetic excitations near the FM zone center ($-$1,1,1) at 10 K and 225 K, respectively. A clear spin gap $E_{g}\approx$ 2.3 meV is observed at 10 K, consistent with the previous report\cite{Liu2020}. The data at 225 K is consistent with a gapless spectrum when the long-range magnetic order disappears. 

\begin{figure}
\centering \includegraphics[width=1\linewidth]{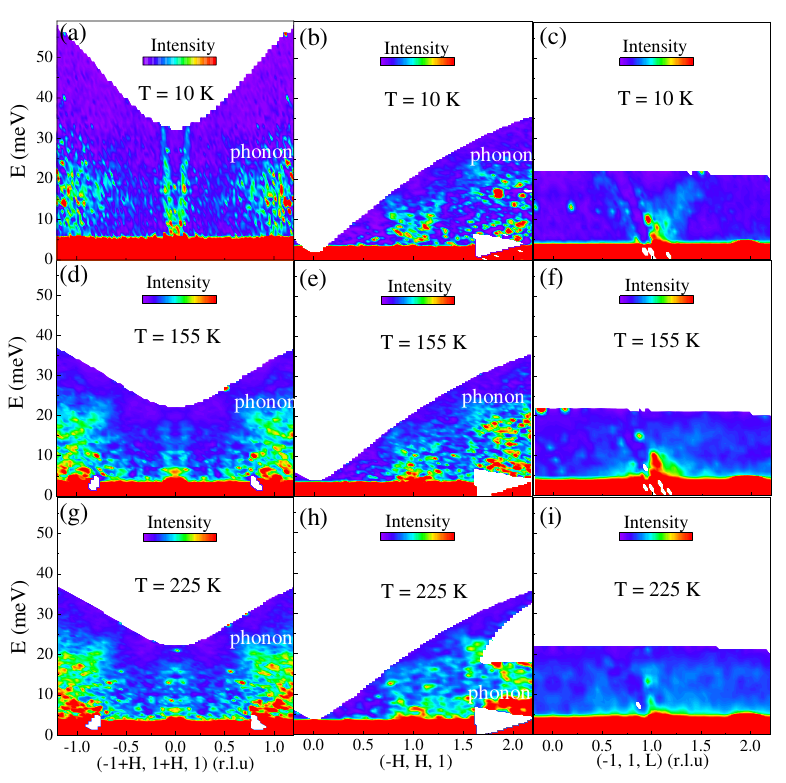} \caption{(color online) 
Magnetic excitations near magnetic zone center (-1 1 1) along [HH0], [$-$HH0] and [0 0 L] direction at (a-c) 10 K, (d-f) 155 K and (g-i) 225 K. Phonons were also observed at higher Q zones. }
\label{fig:Escan} 
\end{figure}

The magnetic excitations along in-plane [HH0], [$-$HH0] and out-of-plane L directions at 5, 155 and 225 K near ($-$1,1,1) are displayed in Fig. \ref{fig:Escan} (a-i). Phonons with energy $\approx$ 18 meV at zone boundaries are also observed in these figures and Fig. S3 (a). At 10 K in the FM state, a remarkable feature is that the dispersion along the [HH0] direction is much steeper yielding a much larger $dE/dQ$ slopes than the [$-$HH0] (or equivalent [H 0 0]) direction, as can also be seen from the very anisotropic excitations in constant-E slice in Fig.~\ref{fig:structure} (f). The dispersion along the L direction shows much smaller slopes than the [HH0] directions too, with $E <$ 30 meV at the zone-boundary ($-$1,1,1.5). Furthermore, the SWs along [HH0] are very sharp and become broader along [$-$HH0]. In addition, significant SW broadening is observed along the L direction. Our results show that Co$_{3}$Sn$_{2}$S$_{2}$ exhibits significant anisotropy in both magnon dispersions and linewidths. Additionally, a strong SW damping is observed as Q approaches the zone boundaries along these three directions (see Fig. S3 in the SM). In $Ref.$\cite{Liu2020}, only the low-E SW dispersions along [H00] and [00L] near the (003) zone were reported below $\approx$ 18 and 15 meV, respectively. Our SW dispersion along the same [00L] direction is similar to that report.  At 155 K, the FM SWs are still visible and become broader without obvious change in the slopes of the dispersions. At 225 K within the paramagnetic state, the dispersions disappear, and evolve into diffuse inelastic signal indicative of the existence of paramagnetic excitations.

 \begin{table*} 
\centering
\setlength{\abovecaptionskip}{0pt}%
\setlength{\belowcaptionskip}{10pt}%
\caption{Summary of the exchange couplings with unit of meV obtained using different approaches. A negative (positive) sign indicates ferromagnetic (antiferromagnetic) coupling. }
\renewcommand{\arraystretch}{1.1}
\begin{tabular}{c|c|c|c|c|c|c|c|c}
\hline\hline
 Approach &$SJ_{1}$ & $SJ_{2}$ & $SJ_{3}$ & S$J_{d}$  & $S Jc_{1}$ & $SJc_{2}$ & $SJc_{3}$ & $Ref.$   \\
 \hline
Fits to SW spectra &   & 5.78  &     &  -29.94 &-11.56   & -1.70  &   & this work     \\
Constrained magnetism & -13.8&   -36.3&  -4.95  &  -4.95 & -31.65   & -13.65  & 15.3  & this work \\ 
MFMST & -7.4118 &    0.1176 &  -1.2941    & -3.6471  & -0.3529 & -5.0588 & -3.5294  & this work  \\ 
MFWF & -13.94 &    -0.9412& -0.1176     & -0.1176   & -1.7647 & -4.4118& -0.6471 & \cite{Liu2020} \\

 \hline\hline  

\end{tabular}

\label{crystal_2}
\end{table*}

\begin{figure}
\centering \includegraphics[width=1\linewidth]{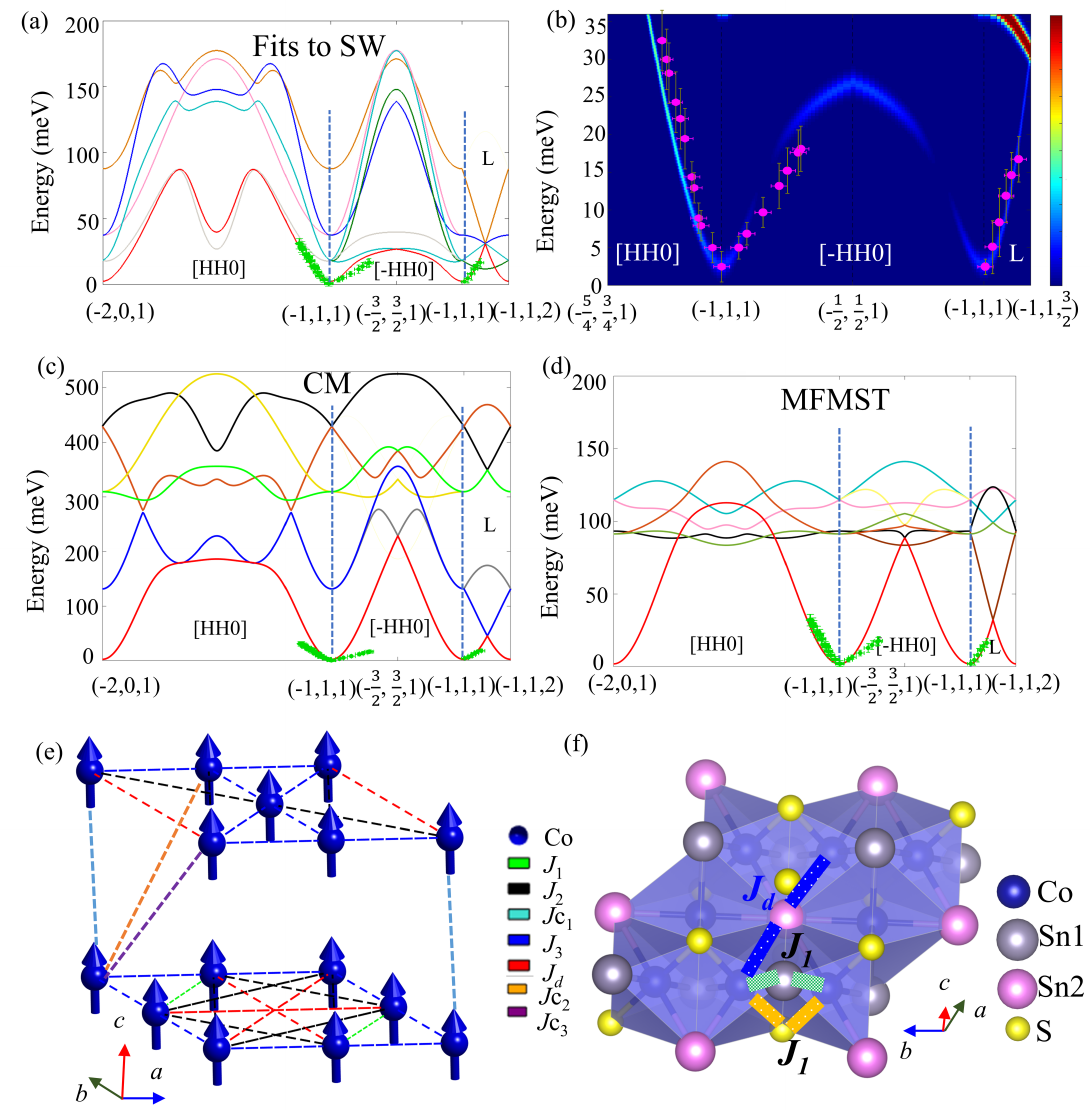} \caption{(color online) Simulation on (a) SW dispersions and (b) intensity after the convolution to the instrument resolution $\approx$ 2 meV using the exchange couplings from the fits to experimental SWs. The dot symbols show the experimental SW dispersions. Simulated SW dispersions using the exchange coupling from (c) constrained magnetism (CM) method and (d) MFMST to illustrate the inability of such calculations to capture the high anisotropy in the experimental SW dispersions. For a better comparison, the same anisotropic term $S A$ of 1.36 was used in all the simulations to reproduce the experimental spin gap. (e). Illustration of Co-Co exchange couplings and the spin configurations. (f). Illustration of compressed octahedral environment of cobalt, geometrical representation of possible exchange pathways Co-S-Co, Co-Sn1-Co for $J_{1}$ and Co-Sn(2)-Co for $J_{d}$.}

\label{fig:SpinWaves} 
\end{figure}

To quantitatively determine the magnetic interactions of the FM ground state in Co$_{3}$Sn$_{2}$S$_{2}$, we have fitted the experimental dispersions and intensities of the SW at 10 K using the SpinW package \cite{Toth2015} for the following spin Hamiltonian: 
\begin{equation}
H=  \sum_{i,j} J_{ij} \mbox{\bf S}_{i} \cdot \mbox{\bf S}_{j}+\sum_{\alpha, \beta, i} S_{i}^{\alpha} A_{i}^{\alpha \beta} S_{i}^{\beta}
\end{equation} where {\bf S}$_i$  is a spin operator, $J_{ij}$ is an exchange coupling between spins, and $A_{i}^{\alpha \beta}$ is a $3\times3$ matrix representing the single-ion anisotropy. The $J_{ij}$ bonds are illustrated in Fig.~\ref{fig:SpinWaves}(e). We have considered in-plane exchange couplings  NN $J_{1}$, next NN (NNN) $J_{2}$, third-neighbor $J_{3}$, inequivalent third-neighbor ``across-hexagon" $J_{d}$ in the kagome lattice, as well as out-of-plane exchange couplings $J_{c1}$, $J_{c2}$ and $J_{c3}$ for both FM and AFM signs in the fits. Extensive effort has been made to consider all the possible models (see the details in the SM). Only the model with dominant FM ``across-hexagon" $J_{d}$ leads to good fits to experimental SW dispersions as shown in Fig.~\ref{fig:SpinWaves}(a).  Although we do not consider the SW damping, the intensities of the simulated SW spectra $S(Q,E)$ shown in Fig.~\ref{fig:SpinWaves}(b) are consistent with the experimental results. The optimized magnetic exchange couplings are summarized in Table I, which includes the strongest FM $J_{d}$, FM $J_{c1}$, FM $J_{c2}$ and very weak AFM $J_{2}$. Attempts to use other models fail to reproduce the very different slopes $dE/dQ$ of the in-plane SW dispersions. Note that the strengths of the third-neighbor $SJ_{3}$ and $SJ_{d}$ are very different and the $J_{1}$ and $J_{3}$ are negligible. The AFM NNN $J_{2}$ exchange is weak but serves to provide frustration with a long-range dominant FM coupling over the diagonal of the kagome hexagons. The single-ion anisotropy term $SA_{i}$ contributes to the emergence of the spin gap.

Experimental spin wave dispersions are used to critique the exchange couplings extracted from the currently popular theoretical approaches. We carried out DFT calculations and confirmed the FM ordering with symmetry R$\mbox{-}$3m' is most stable as observed experimentally. The ordered Co moment is found to be $\approx 0.35 \mu_B$ ($S\approx0.17$), well consistent with the experimental results. We then examine exchange couplings theoretically. First we employ the constrained magnetism method  and find $SJ_{2}$ is the strongest one. Second, we use a magnetic force theorem within multiple scattering theory (MFMST)\cite{Liechtenstein1987,Ebert2009} and the resultant exchange couplings belong to dominant $J_{1}$ model. The magnetic exchange couplings were also reported in Ref.~\cite {Liu2020} by employing a combination of the magnetic force theorem \cite{Liechtenstein1987} and Wannier functions approach (MFWF)\cite{Korotin2015}, which belongs to the dominant $J_{1}$ model too. The magnetic exchange couplings obtained from these three theoretical methods are summarized in Table I. 

The simulated SW dispersions using these exchange couplings are displayed in Fig.~\ref{fig:SpinWaves}(c-d) and Fig. S4, respectively. The theoretical SW dispersions significantly deviate from the experimental results, indicating these three methods have shortcomings for extracting the reliable exchange couplings in Co$_{3}$Sn$_{2}$S$_{2}$. The main reason is that Co$_{3}$Sn$_{2}$S$_{2}$ is a weakly-correlated itinerant system with small moment in which the size of magnetic moment depends on its deviation from ground state. However, DFT calculations use rigid spin approximation, i.e., moment size independent from its orientation. Constrained magnetism calculations overestimate the exchange coupling because these calculations necessarily force ordered moments for metastable magnetically ordered states, and therefore overestimate the energy of such states. The MFMST or MFWF methods calculate the energy change due to small magnetic moment deviation from the ground state direction under the assumption that the moment size does not change and may underestimate the magnetic exchange couplings as in fcc Ni\cite{Buczek2011,Schilfgaarde1999}.

The dominant third-neighbor $S J_{d}$ across the hexagons is very unusual in kagome lattice systems. To shed light on it, we examine the crystal structure and discuss the possible exchange pathways. As illustrated in Fig.~\ref{fig:SpinWaves}(f), the kagome lattice is formed by the cobalt atoms in the $ab$ layer and centered by Sn(2) atoms. The cobalt atoms are octahedrally coordinated by two axial sulfur atoms and four tin atoms, i.e., two in-equivalent Sn(1) and Sn(2), with shared faces among them within the $ab$ plane. Interestingly, the cobalt-centered octahedra are strongly compressed, with much shorter Co-S distance 2.1729 \AA{} than  Co-Sn(1) distance 2.6779 \AA{} and Co-Sn(2) one 2.6796 \AA{} at 10 K. There are two possible exchange pathways lying above/below the kagome ($ab$) layers: Co-S-Co with the bond angle of 76.14$^\circ$ and  Co-Sn1-Co with a bond angle of 60.04$^\circ$. The large difference in the bond lengths and angles may result in two competing AFM and FM magnetic interactions that cancel out, leading to a negligible $J_{1}$ like in BaCu$_{3}$V$_{2}$O$_{8}$(OD)$_{2}$ \cite{Boldrin2018}.
Because of the shortest distance between Co and S, they form hybridized bands, which are dominated by Co near the Fermi level, i.e., antibonding bands (see SM for more details).  
While the Co-Sn distance is longer than the Co-S distance, Sn states are also found to contribute to the electronic bands near the Fermi level, along the $\Gamma$-T line and near the L point. 
The former is expected to contribute to the in-plane strongest FM $J_d$ coupling via Co-Sn(2)-Co across the Co hexagon since $k_x=k_y=0$, and the latter is expected to contributes to the out-of-plane exchange couplings $J_{c1}$and $J_{c2}$ because of non-zero momentum.

\begin{figure}
\begin{center}
\includegraphics[width=1\columnwidth, clip]{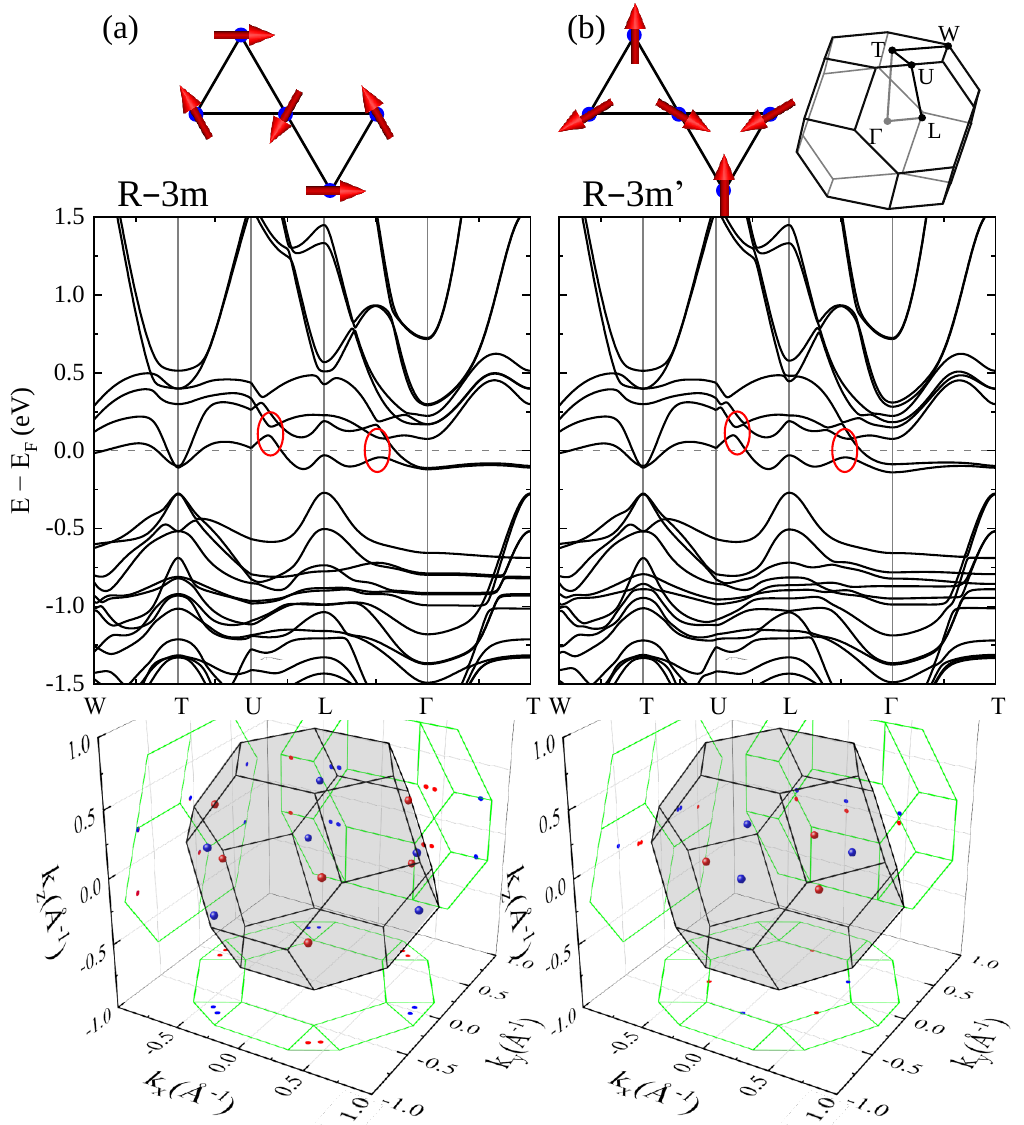}
\caption{Spin configuration (top panel), band structure (middle panel) and Weyl points in the first Brillouin zone (bottom panel) for (a) the $\rm R\mbox{-}3m$ AFM ordering and (b) the R$\mbox{-}$3m' AFM ordering. Right top figure shows the first Brillouin zone with high symmetry points.
Red circles indicate the reminiscent of nodal lings in the FM state without SOC. For the $\rm R\mbox{-}3m$ AFM, only Weyl points near zone boundaries are plotted. Blue (red) points in the bottom panel indicate chirality  $+1 (-1)$. }
\label{fig:dispersion}
\end{center}
\end{figure}
 
We next examine the electronic band structure in the ground state and metastable magnetic states using the constrained magnetism method of DFT. Although this method overestimated the magnetic exchange couplings, it is expected to provide reliable electronic band structures because these are single-particle properties governed by the underlying symmetry and the current system is weakly correlated.  Our electronic band structure for the FM ordering (see Fig. S7) is indeed well consistent with the previous theoretical results \cite{Liu2018} and experimental ARPES measurements \cite{Liu2018}.
The previous work proposed  $\rm R\mbox{-}3m$ 120$^\circ$ AFM ordering \cite{Guguchia2020} in $T_{A}<T<T_{C}$. In fact, there are two symmetry inequivalent AFM structures of $\rm R\mbox{-}3m$ and R$\mbox{-}$3m' (same symmetry as FM order) as schematically shown in Fig.~\ref{fig:dispersion}~(a) and (b), respectively. We found $\rm R\mbox{-}3m$ has lower energy than R$\mbox{-}$3m' by $\sim 1$~meV per formula unit. Since the difference in energy is rather small, this suggest R$\mbox{-}$3m' structure may  be also considered in analyzing the electronic properties in this temperature region. 

The electronic band structures for the two AFM states are shown in Fig.~\ref{fig:dispersion}. In contrast to the band structure of the FM ground state, there appear two electron pockets centered at the T point, where the two bands nearly touch. Despite their distinct band structures than that of FM order, the two AFM states have a similar feature, where two lowest conduction bands get closer along the U-L line and the L-$\Gamma$ line (see red circles in Fig.~\ref{fig:dispersion}), reminiscent of the nodal ring appearing in the FM state without the SOC. In fact, the R$\mbox{-}$3m' AFM state has the combined time-reversal $\cal T$ and mirror-reflection ${\cal M}_y$ symmetry. This symmetry protects 6 Weyl points in the plane running on U-L-$\Gamma$ points shown in the right bottom figure in Fig.~\ref{fig:dispersion}, as in the FM state (see detailed analysis in the SM).
We also found that the $\rm R\mbox{-}3m$ AFM state has Weyl points. 
However, the locations of Weyl points are different because the combined $\cal T$-${\cal M}_y$ symmetry is broken in $\rm R\mbox{-}3m$  AFM. 
Most of these Weyl points appear near the $\Gamma$-T line, where the band structure shows nearly degenerate bands $\sim 0.1$~eV below the Fermi level. Another 12 Weyl points are located near zone boundaries (see the left bottom figure in Fig.~\ref{fig:dispersion}). 
In both AFM states, Weyl points do not contribute to the anomalous Hall conductivity because of the twofold rotational symmetry ${\cal C}_2$ with the rotation axes lying in the $ab$ plane \cite{Guguchia2020}. Thus, we conclude that there probably exists a different Weyl state in $T_{A}<T<T_{C}$ no matter which type of 120$^\circ$ AFM order coexists with the FM order.

In summary, we report the anisotropic magnetic excitations, unusual exchange couplings and  correlations between various magnetic orders and the electronic band topology in the Weyl semimetal Co$_{3}$Sn$_{2}$S$_{2}$.  In an intermediate temperature region $T_{A}<T<T_{C}$, a new Weyl state supported by the 120$^\circ$ AFM order is predicted. ARPES would be very useful to detect the Weyl points in this temperature region and to distinguish two possible 120$^\circ$ AFM orders. Below $T_{A}$, the simple FM order is dominated by FM third-neighbor ``across-hexagon" coupling with a weak frustrated NNN bond. We further demonstrate the rigid spin approximation based calculations have shortcomings for extracting reliable magnetic exchange couplings in Co$_{3}$Sn$_{2}$S$_{2}$ and it requires advances in the theory of spin-spin interactions that do not rely on the existence of rigid local moments, for instance direct calculation of dynamic magnetic susceptibility from weak-coupling approaches\cite{Buczek2011,Izuyama1963,Savrasov1998}.

\begin{acknowledgments}
 We would like to thank Dr. S. Mankovsky for fruitful discussions. The research by SO, JY, MAM, DAT was sponsored by the Laboratory Directed Research and Development Program (LDRD) of Oak Ridge National Laboratory, managed by UT-Battelle, LLC, for the U.S. Department of Energy (Project ID 9533), with later stage supported by the US DOE, Office of Science, Basic Energy Sciences, Materials Sciences and Engineering Division (JY), and U.S. Department of Energy, Office of Science, National Quantum Information Science Research Centers, Quantum Science Center (SO, MAM, DAT). The neutron research used resources at Spallation Neutron Source, a DOE Office of Science User Facility operated by the Oak Ridge National Laboratory. GDS was sponsored by the Laboratory Directed Research and Development Program of Oak Ridge National Laboratory, managed by UT-Battelle, LLC, for the U. S. Department of Energy. DM and RX acknowledge support from the Gordon and Betty Moore Foundation's EPiQS Initiative, Grant GBMF9069.\\
 Copyright  notice: This  manuscript  has  been  authored  by  UT-Battelle, LLC under Contract No. DE-AC05-00OR22725 with the U.S.  Department  of  Energy.   
The  United  States  Government  retains  and  the  publisher,  by  accepting  the  article  for  publication, 
acknowledges  that  the  United  States  Government  retains  a  non-exclusive, paid-up, irrevocable, world-wide license to publish or reproduce the published form of this manuscript, 
or allow others to do so, for United States Government purposes.  
The Department of Energy will provide public access to these results of federally sponsored  research  in  accordance  with  the  DOE  Public  Access  Plan 
(http://energy.gov/downloads/doe-public-access-plan)
\end{acknowledgments}


\end{document}


\title{Supplemental Material for \\ \hspace*{\fill} \\
Unusual Exchange Couplings and Intermediate Temperature Weyl State in Co$_{3}$Sn$_{2}$S$_{2}$ }

\author{Qiang Zhang}

\affiliation{Neutron Scattering Division, Oak Ridge National Laboratory, Oak Ridge, Tennessee 37831, USA}

\author{Satoshi Okamoto}
\affiliation{Materials Science and Technology Division, Oak Ridge National Laboratory, Oak Ridge, Tennessee 37831, USA}

\author{German D. Samolyuk}
\affiliation{Materials Science and Technology Division, Oak Ridge National Laboratory, Oak Ridge, Tennessee 37831, USA}

\author{Matthew B. Stone}
\affiliation{Neutron Scattering Division, Oak Ridge National Laboratory, Oak Ridge, Tennessee 37831, USA}

\author{Alexander I. Kolesnikov}
\affiliation{Neutron Scattering Division, Oak Ridge National Laboratory, Oak Ridge, Tennessee 37831, USA}

\author{Rui Xue}
\affiliation{Department of Physics $\&$ Astronomy, The University of Tennessee, Knoxville, TN 37996, USA}

\author{Jiaqiang Yan}
\affiliation{Materials Science and Technology Division, Oak Ridge National Laboratory, Oak Ridge, Tennessee 37831, USA}

\author{Michael A. McGuire}
\affiliation{Materials Science and Technology Division, Oak Ridge National Laboratory, Oak Ridge, Tennessee 37831, USA}
\affiliation{Quantum Science Center, Oak Ridge, Tennessee 37831, USA}

\author{David Mandrus}

\affiliation{Department of Materials Science and Engineering, University of Tennessee, Knoxville, TN 37996, USA}
\affiliation{Materials Science and Technology Division, Oak Ridge National Laboratory, Oak Ridge, Tennessee 37831, USA}
\affiliation{Department of Physics $\&$ Astronomy, The University of Tennessee, Knoxville, TN 37996, USA}

\author{D. Alan Tennant}
\affiliation{Materials Science and Technology Division, Oak Ridge National Laboratory, Oak Ridge, Tennessee 37831, USA}
\affiliation{Shull Wollan Center, Oak Ridge National Laboratory, Oak Ridge, Tennessee 37831, USA}
\affiliation{Quantum Science Center, Oak Ridge, Tennessee 37831, USA}

\date{\today}

 \maketitle

 \textbf{\uppercase\expandafter{\romannumeral1.} Preparation of polycrystalline and single crystal samples}
 
   To synthesize polycrystalline  Co$_{3}$Sn$_{2}$S$_{2}$, first Co slugs and Sn shot in a 3:2 molar ratio were sealed in an evacuated silica ampoule that was then heated to 1150 $^\circ$C and held for several hours. This produced single phase Co$_{3}$Sn$_{2}$, which has a melting point near 1200$^\circ$C. This was then ground into a coarse powder and combined with sulfur pieces in a 1:2 molar ratio. These reactants were sealed in an evacuated silica ampoule, heated to 900$^\circ$C, and held for 2 days. This generally produced single phase Co$_{3}$Sn$_{2}$S$_{2}$, as judged by powder x-ray diffraction. In some batches, binary phases due to incomplete reactions were observed at the level of a few percent. These impurities could be reduced or eliminated by grinding and repeating the 900$^\circ$C heat treatment. Single phase Co$_{3}$Sn$_{2}$S$_{2}$ powder ($\sim$ 6 g) was then sealed under vacuum in a carbon coated quartz tube with a tapered end.  The best crystals were obtained using a tube with an outer diameter of 12mm and a wall thickness of 1.5mm. The growth was performed in a vertical Bridgman furnace \cite{May2020} in Oak Ridge National Laboratory. The growth ampoule was first heated to 1020$^\circ$C at 180$^\circ$C/h. After dwelling at this temperature for 12 hours, it is cooled down to 860$^\circ$C at 2$^\circ$C/h. The furnace was then turned off after homogenizing at 860$^\circ$C for 24 h. The as-grown ingot normally has 3-4 large grains and the grain boundary can be easily observed with the help of an optical microscope. The single crystal with mass $\approx$ 1.2 g used in the inelastic neutron scattering was cut from such a large ingot.
  
\begin{figure}
\begin{center}
\includegraphics[width=1\columnwidth, clip]{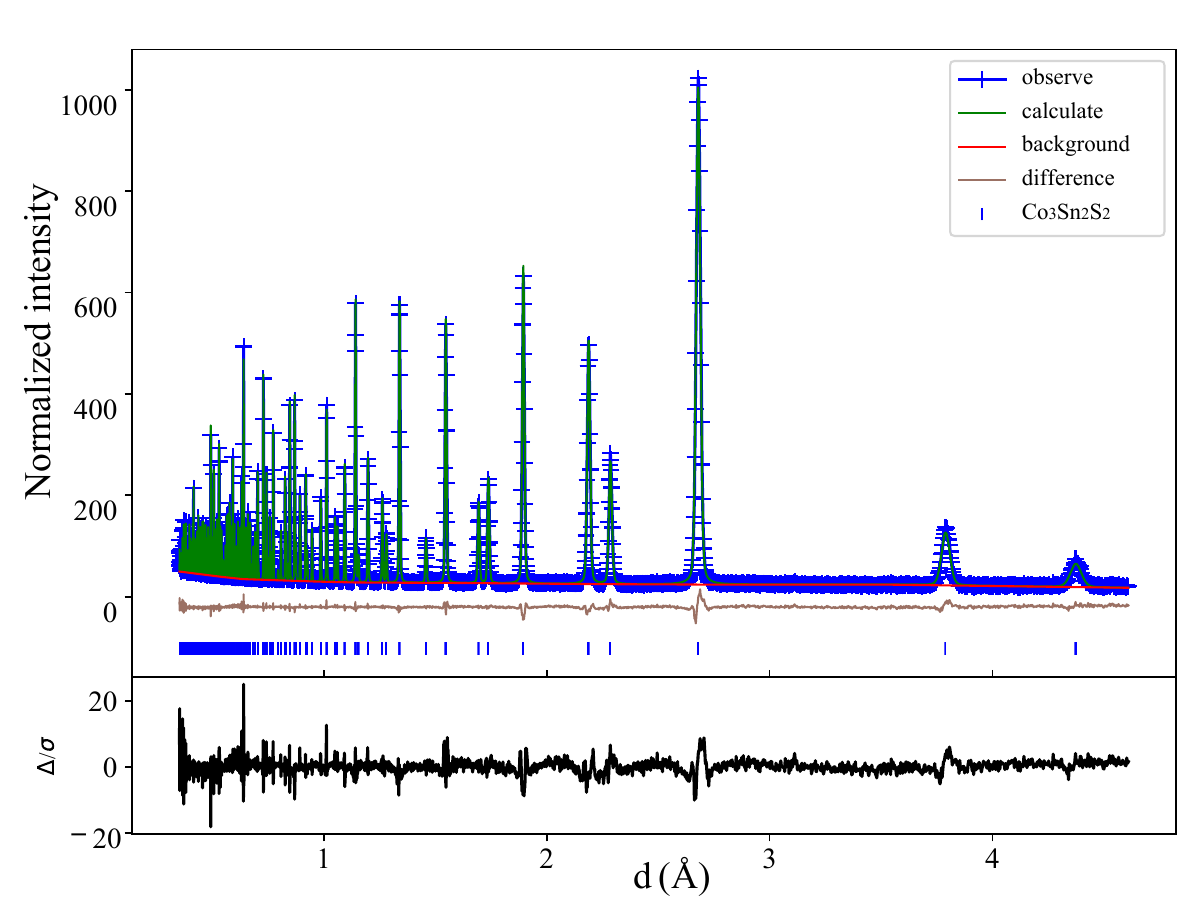}
\caption{Rietveld analysis on the powder neutron patterns at 10 K}
\label{fig:NPD}
\end{center}
\end{figure}

\begin{table} 
\centering
\setlength{\abovecaptionskip}{10pt}%
\setlength{\belowcaptionskip}{10pt}%
\caption{Refined atomic positions and isotropic temperature factors determined from modeling powder neutron diffraction data of Co$_{3}$Sn$_{2}$S$_{2}$ at 10 K. Analysis of the nuclear Bragg reflections lead to the space group
selection of $\rm R\mbox{-}3m$ (No. 166)- and indexed unit cell constants of $a = b=$ 5.359(3)  \AA{},  $c =$
13.115(2) \AA{}.}
\renewcommand{\arraystretch}{0.8}
\begin{tabular}{c|c|c|c|c|c}
\hline\hline
 atom & Wyckoff Site&  x & y &  z& $ ^{a}U_{eq}$    \\
 \hline
Co & 9d &    0.5&  0     &   0.5  & 0.00051(5)   \\
Sn1& 3a &    0&  0     &   0   &  0.00076(6)  \\ 
Sn2 & 3b &    0&  0     &   0.5  &   0.00076(6)  \\
 S & 6c &    0&  0     &   0.283(4)  &  0.00165(7)   \\
 \hline\hline  

\end{tabular}
\label{crystal_2}
\end{table}

 \textbf{\uppercase\expandafter{\romannumeral2.} Neutron powder diffraction measurement}

 Neutron diffraction measurements at 10 K were conducted on the polycrystalline sample at the high-resolution Time-of-flight(TOF) powder diffractometer POWGEN, located at SNS, ORNL. Around 10 g of powder was loaded in a vanadium can with helium exchange gas. The incident neutron beam with center wavelength of 0.8 \AA{} was used to get a large $d$ coverage of 0.2 \AA{}- 8 \AA{}. Rietveld analysis on the high-resolution neutron pattern shows that the rhombohedral structure with space group selection of $\rm R\mbox{-}3m$ (No. 166) can fit the data very well, as shown in Fig.~\ref{fig:NPD}. There is no indication of the lattice distortion or lower symmetry of the crystal structure. We did not find a clear ferromagnetic (FM) signal manifested as the increase of the peak intensity on allowed nuclear peak positions. This is ascribed to the very weak magnetic moment $\approx$0.37 $\mu_{B}$ that is below the detectable moment $\approx$ 0.4 $\mu_{B}$ at POWGEN. The refined lattice parameters, atomic positions and isotropic temperature factors  at 10 K of  Co$_{3}$Sn$_{2}$S$_{2}$ are summarized in Table SI. Such accurate structural information is used as the input for DFT calculations.

\begin{figure}
\begin{center}
\includegraphics[width=1\columnwidth, clip]{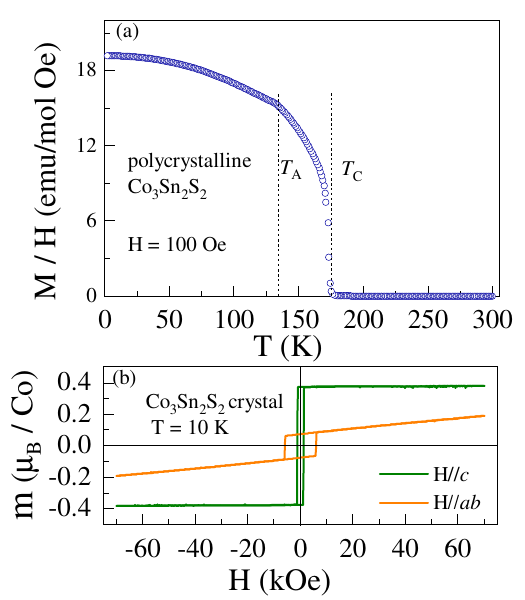}
\caption{ (a). Temperature dependence of susceptibility of polycrystalline Co$_{3}$Sn$_{2}$S$_{2}$. (b). Magnetic hysteresis loops of Co$_{3}$Sn$_{2}$S$_{2}$ crystal with field parallel to $c$ axis and $ab$ plane.}
\label{fig:susceptibility}
\end{center}
\end{figure} 

\textbf{\uppercase\expandafter{\romannumeral3.} Magnetic characterizations}

\begin{figure}
\begin{center}
\includegraphics[width=1\columnwidth, clip]{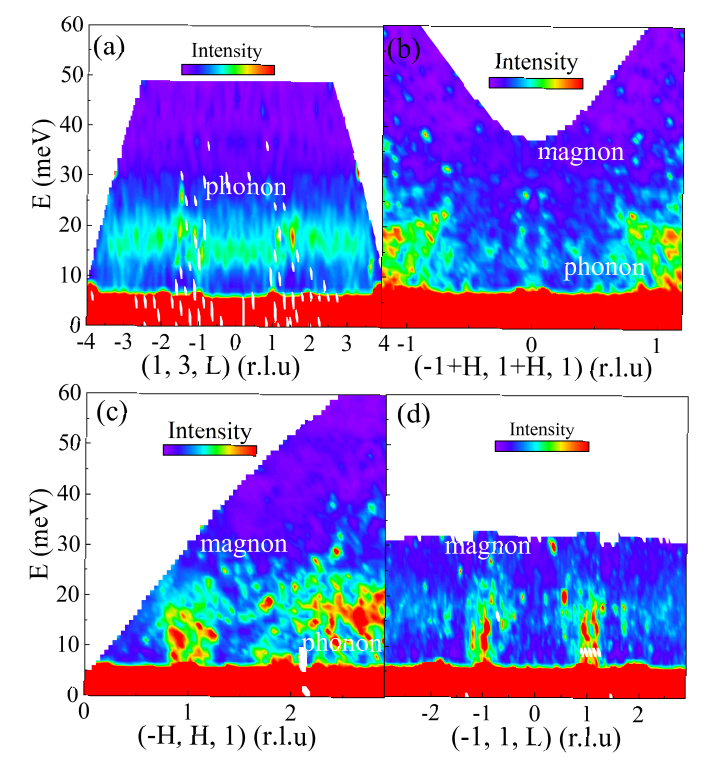}
\caption{Observation of (a) phonons at high-Q (1, 3, L) zones and the spin waves damping using (b) E$_{i}$= 100 meV along [HH0], (c) E$_{i}$=75 meV along [-HH0] and (d) E$_{i}$=75 meV along L direction at 10 K. }
\label{fig:INS}
\end{center}
\end{figure}

X-ray Laue diffraction was used to orient the single crystals with an accuracy of 1 degree. The anisotropic magnetic properties with magnetic field parallel to the $ab$ plane and the $c$ axis were measured with a Quantum Design MPMS3 SQUID magnetometer in the temperature range 2 K$\leq T \leq $300 K. The temperature dependences of the susceptibility in Fig. 1(c) for the single crystal and Fig.~\ref{fig:susceptibility}(a) for polycrystalline sample were collected using a field-cooled mode in an applied magnetic field of 250 Oe and 100 Oe, respectively.

\textbf{\uppercase\expandafter{\romannumeral4.}Single-crystal inelastic neutron scattering experiments and fits to spin waves using SpinW package} 

 Inelastic neutron scattering experiments were performed using the  SEQUOIA spectrometer located also at SNS. The crystal was aligned in the (H K 0) horizontal scattering plane with a mosaicity of $\sim2$ degree. The data were collected at 5, 155 and 225 K using incident energies of 15, 40, 75 and 100 meV. The constant-energy ($E$) or Q cuts were fitted using a Lorentzian function to obtain both the spin wave (SW) dispersion and intensity. The fits to them using the SpinW package \cite{Toth2015,Zhang2019} based on linear spin wave theory yield the magnetic exchange constants and single-ion anisotropy.
 \begin{figure}
\begin{center}
\includegraphics[width=1\columnwidth, clip]{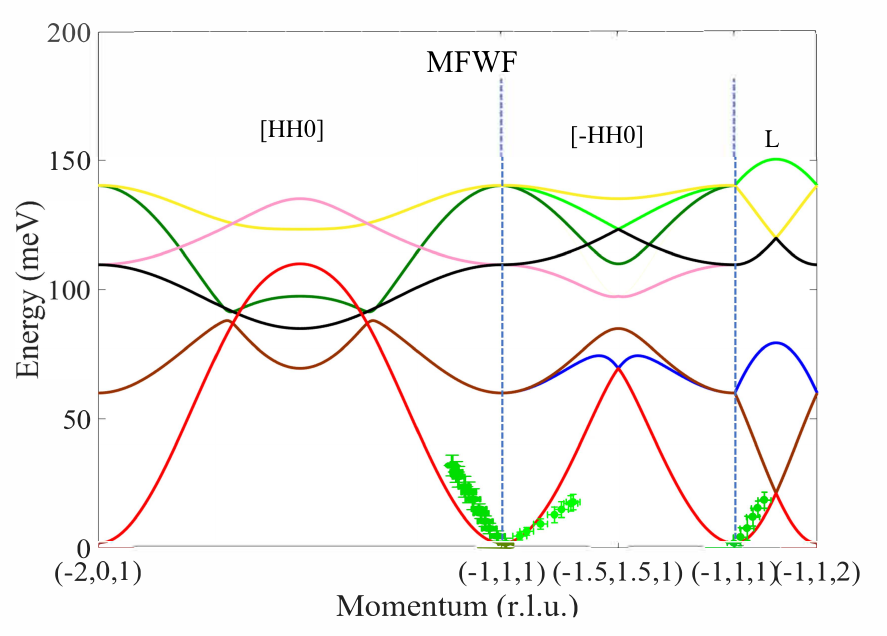}
\caption{ Simulation on spin wave dispersions using the exchange couplings from Ref. \cite{Liu2020} by employing a combination of the magnetic force theorem \cite{Liechtenstein1987} and Wannier functions approach (MFWF). The dot symbols show the experimental SW dispersions. }
\label{fig:MFWF}
\end{center}
\end{figure}
    
  \begin{figure*}
\centering \includegraphics[width=1\linewidth]{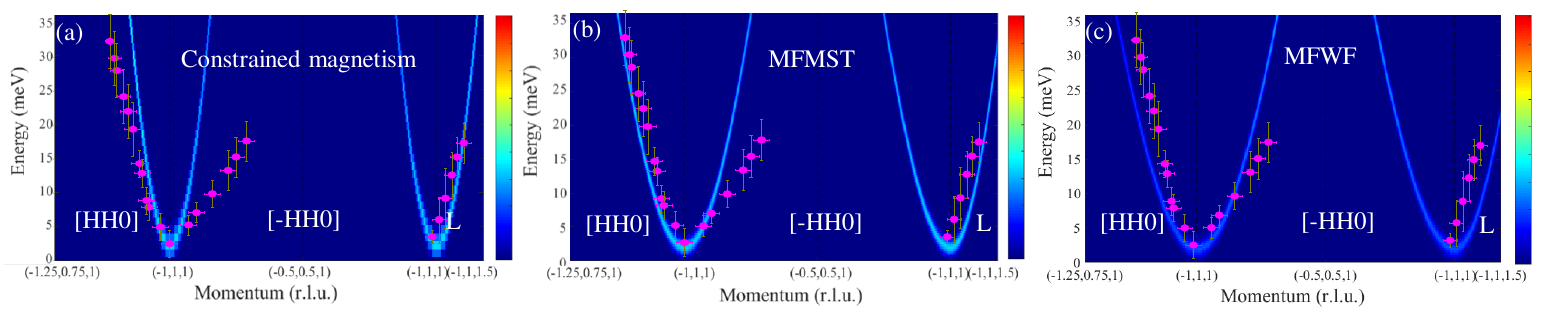} \caption{Simulations of low-E SW spectra using (a) constrained magnetism, (b) the magnetic force theorem within multiple scattering theory (MFMST) and (c) MFWF in Ref. \cite{Liu2020}. The dot symbols show the experimental SW dispersions. }
\label{fig:LowESW} 
\end{figure*}

As shown in Fig. 2 (a-c) and 1 (f), the dispersion along the [HH0] direction is much steeper yielding a much larger dE/dQ slopes than the [-H H 0], which is unusual for the Kagome lattice. We have  experimental SW dispersions up to q$_{h}$ =0.13 in [-1+ q$_{h}$, 1+q$_{h}$, 1], q$_{k}$ =0.3 in [-1- q$_{k}$, 1+ q$_{k}$, 1] and q$_{l}$=0.35 in [-1- q$_{l}$, 1+ q$_{l}$, 1]. Along the [HH0] direction, our INS measurements show that the energy at q$_{h}$=0.13 is as high as 32 meV. In sharp contrast, along [-HH0] direction, the energy at q$_{k}$=0.3 (close to the zone boundary  q$_{k}$=0.5) is as low as $\approx$18meV. Along L direction,  the energy at q$_{l}$=0.35 (also close to the zone boundary  q$_{l}$=0.5) is as low as $\approx$17meV. Thus, although INS measurements could not map out the whole dispersions along these three directions, the spin wave fits to our availabe experimental SW dispersions are sufficient to distinguish different models. Extensive effort has been made to consider all the possible models including but not limited to: 1) the strongest $Jc_{1}$, $Jc_{2}$ and/or $Jc_{3}$ model; 2). the strongest NN $J_{1}$ model; 3). the strongest NNN $J_{2}$ model;  4). competing in-plane AFM and FM exchange couplings with the combination to out-of-plane $Jc_{1}$, $Jc_{2}$ and/or $Jc_{3}$ model; 5). the strongest third-neighbor $J_{d}$ or $J_{3}$ model;  5). the strongest 4th-neighbor $J_{4}$ model. Note that we considered both AFM and FM signs for all the exchange couplings in the fits.  Only the model with dominant FM ``across-hexagon" $J_{d}$ leads to good fits to experimental SW dispersions as shown in Fig. 3(a-b). All the other models cannot fit the experimental dispersions along three directions simultaneously, especially the very anisotropic dispersions with very different dE/dQ slopes along the [HH0] and [-HH0] directions.

\begin{figure}
\begin{center}
\centering \includegraphics[width=1\linewidth]{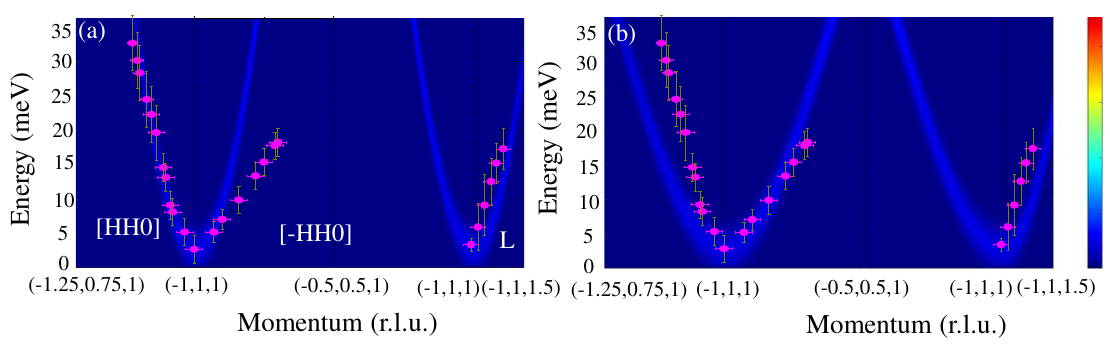}
\caption{ The simulated spectra after optimizing the exchange couplings on the MFWF model to get good fits to (1) [ H H 0] or (2) [-HH0] direction. }
\label{fig:TwoCases}
\end{center}
\end{figure}

 For example, the comparison between the experimental SW dispersions to the simulated dispersions using the three theoretical methods of constrained magnetism, MFMST, MFWF are displayed in Fig. 3(c-d) and Fig.~\ref{fig:MFWF}, respectively. The comparisons of experimental SW dispersion and simulated SW spectra $S(Q, E)$ in low E scale are shown in  Fig.~\ref{fig:LowESW}(a-c).
  These theoretical methods cannot reproduce the experimental results. Please note that we have also used such theoretical methods to improve the quality of the fits by optimizing the exchange couplings. But we still failed to get reasonable fits. Let us take the MFWF method that belongs to dominant $J_{1}$ model  reported in Ref.\cite{Liu2020} for demonstration. As compared to the experimental SW dispersions (see Fig.~\ref{fig:MFWF} and Fig.~\ref{fig:LowESW}(c)), the dE/dQ slopes along the [HH0] direction in the simulated SW dispersion for the MFWF model are smaller but the dE/dQ slopes along [-HH0] direction are much larger beyond q$_{k}\approx$ 0.16. If we increase dE/dQ slopes to get good fits to the dispersion along [HH0] by adjusting the exchange couplings, the slopes along [-HH0] direction become even higher as shown in the Fig.~\ref{fig:TwoCases}(a). On the contrary, if we fit the experimental dispersion along [-HH0] well, the slopes along [HH0] becomes far smaller than the experimental dispersion as displayed in Fig.~\ref{fig:TwoCases}(b). It is unlikely to get reasonable fits to very anisotropic dispersions with very different dE/dQ slopes along the [HH0] and  [-HH0] directions using a dominant $J_{1}$ model.

  It turns out that any of the strongest $J_{1}$, $J_{2}$, $J_{3}$, $Jc_{1}$ or $Jc_{2}$ model yields a much smaller difference between the dE/dQ slopes along the [HH0] and [-HH0] directions than that in the experimental dispersions. We found it is robust that only the dominant $J_{d}$ model could reproduce the very anisotropic dispersions along these two directions. 

\textbf{\uppercase\expandafter{\romannumeral6.} Details of multiple-scattering Green function calculations for exchange coupling}

 We also examined magnetic exchange coupling using linear response approach (utilizing force theorem) \cite{Liechtenstein1987} within relativistic multiple-scattering formalism \cite{Ebert2009} as it implemented in SPR-KKR package  \cite{Ebert2011,Ebert2017}. Here the electronic structure was calculated with the angular momentum cutoff lmax$=$3 and exchange-correlation energy parametrization by Perdew, Burke, and Ernzerhof (PBE) \cite{Perdew1996}. Correlation on Co atoms were incorporated within the generalized gradient approximation $+U$, GGA$+U$ \cite{Liechtenstein1995} approach with $U=$1.1 and $J=$0.1 eV. The improvement of magnetic properties description by using GGA+$U$ approach with small U values was demonstrated in a series of works including {Co}. In order to improve the accuracy of the multiple-scattering approach for the case of ``open'' structure Co$_{3}$Sn$_{2}$S$_{2}$ three empty spheres with radius 1.3 Bohr were introduced. Integration over the Brillouin zone was executed over 5000 k-points, which guarantees accuracy of a few percents for exchange coupling calculation.
 
\textbf{\uppercase\expandafter{\romannumeral5.} Details of DFT calculations for magnetic ordering and electronic band structure}

We performed DFT calculations using standard methods on Co$_{3}$Sn$_{2}$S$_{2}$. We use the accurate structural information at 10~K obtained from our powder neutron diffraction measurements as the input. The projector augmented wave method \cite{Kressee1999} is used with the generalized gradient approximation in 
the parametrizeation of Perdew, Burke, and Enzerhof \cite{Perdew1996} for exchange-correlation as implemented in the Vienna {\it ab initio} simulation package (VASP) \cite{Kressee1996}. For Co and S standard potentials are used (Fe and S in the VASP distribution), 
while for Sn a potential in which $d$ states are treated as valence states, is used (Sn$_d$).  In most cases, we use an $8 \times 8 \times 4$ $k$-point grid and an energy cutoff of 500~eV. The spin-orbit coupling (SOC) is included, but the $+U$ correction is not included 
because Co$_3$Sn$_2$S$_2$ is an itinerant magnetic system.

\begin{figure}
\begin{center}
\includegraphics[width=0.8\columnwidth, clip]{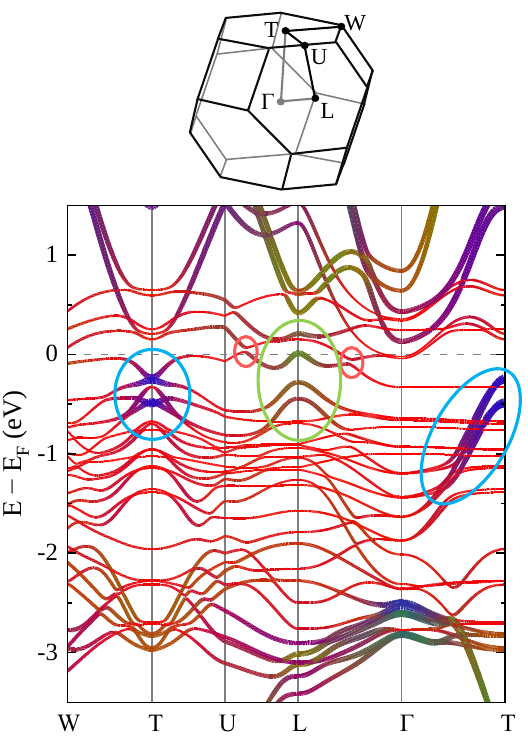}
\caption{Band structure for the ferromagnetic ordered state with magnetic moments pointing to the $c$ axis (FM) with the SOC turned on. 
The line width is proportional to the total weight of Sn, with the weight of Sn(1) is indicated by green, Sn(2) by blue, and Co by red. 
Light red circles indicate the location of a nodal ring gapped by the SOC.
A light green circle indicates the finite contribution of Sn(1) to the electronic bands near the Femi-level, while light blue circles indicate the finite contribution of Sn(2). These indicate the contribution of Sn to magnetic exchange coupling. 
The first Brillouin zone is shown on the top.}
\label{fig:fatband}
\end{center}
\end{figure}

One way to evaluate the magnetic exchange coupling using DFT is calculating the total energy of 
different magnetically ordered states and map their energy to the Heisenberg form. 
In Co$_3$Sn$_2$S$_2$, we realized that all the antiferromagnetic (AFM) states considered are unstable in DFT. 
Therefore, to evaluate the magnetic exchange coupling using DFT, 
we carry out the constrained magnetism calculations considering various AFM states by setting ${\rm I\_CONSTRAINED\_M}=2$. We gradually increase the value of $\rm LAMBDA$ so that the penalty to the total energy $\rm E\_p$ becomes sufficiently small. In some cases, we increase the size of the unit cell to examine long-range exchange coupling. By mapping the total energy of these magnetic orderings onto the Heisenberg model with the spin anisotropy, we obtained the magnetic exchange couplings. The $S$ in the ground state, where Co spins are ordered ferromagneically along the $c$ axis (FM), is found to be $\approx 0.17$, consistent with the our experimental result with ordered moment 0.37 $\mu_B$ ($m=gS$, $g\approx$2). The uniaxial spin anisotropy $S A$ is determined to be 1.33, consistent with 1.36 from SW fits. However, the simulated SW dispersions using these exchange couplings are inconsistent with the experimental SW dispersions as shown in the Fig. 3(c) and Fig.~\ref{fig:LowESW}(a).

\begin{figure}
\begin{center}
\centering \includegraphics[width=1\linewidth]{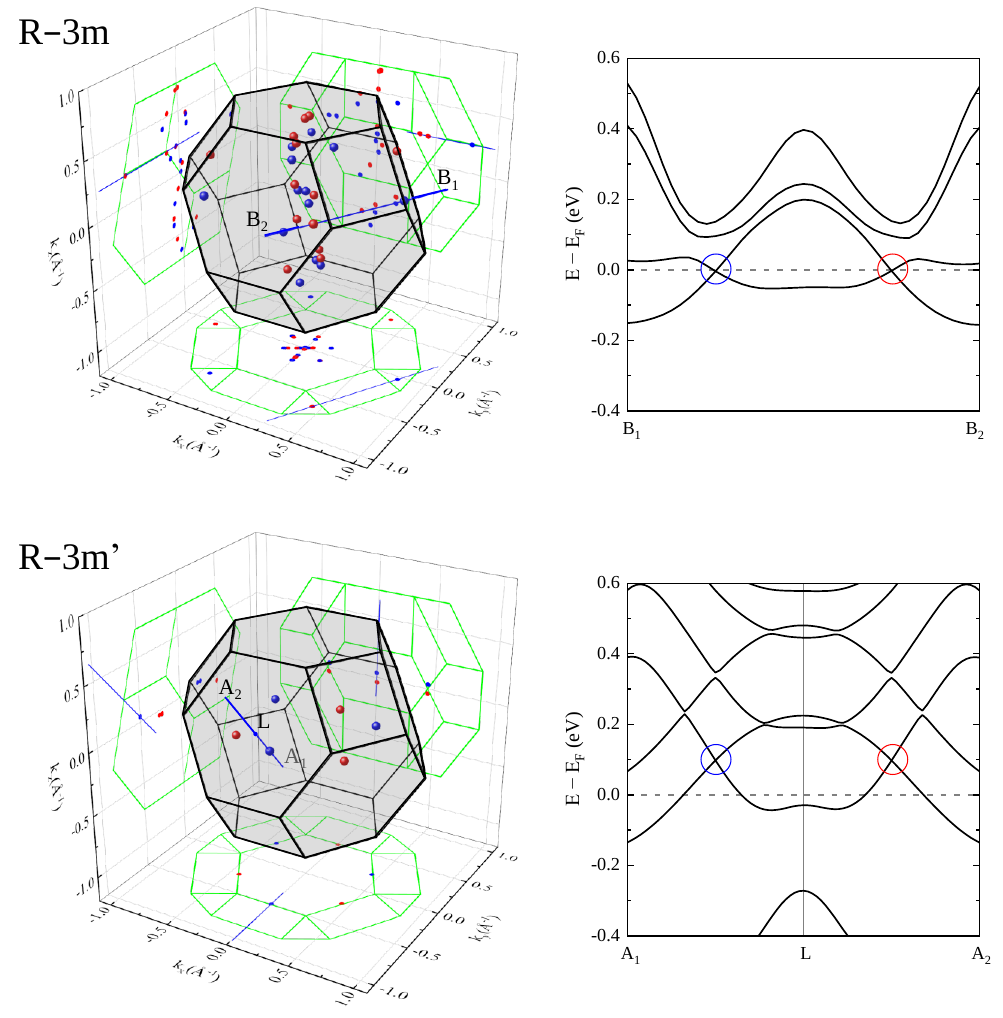}
\caption{Comprehensive Weyl points in R$-3$m (top) and R$\mbox{-}$3m' (bottom) 120$^\circ$ AFM states near both $kx=ky=0$ and the zone boundaries in the first Brillouin zone. 
Blue (red) points indicate chirality  $+1 (-1)$. 
Right panels show the band dispersions along blues lines in the left panels.
Weyl points with chirality  $+1 (-1)$ are indicated by blue (red) circles. }
\label{fig:weyl}
\end{center}
\end{figure}
 
 Figure \ref{fig:fatband} shows the dispersion relation in the FM state. 
The energy vs. momentum relation is fully consistent with the previous report \cite{Liu2018}. 
Here, the line width is proportional to the total weight of Sn with the color code indicating individual contributions; the weight of Sn(1) is indicated by green, Sn(2) by blue, and Co by red. 
The electronic bands near the Fermi level are dominated by Co hybridized with Sn and S. 
The strong contribution from S  is found to appear away from the Fermi level, $E-E_F \alt -4$~eV (not shown). 
On the other hand, the Sn contribution appears near the Fermi level. 
As indicated by light blue circles, Sn(2) has the large weight along the $\Gamma$-T direction, i.e., $k_x=k_y=0$. 
Since Sn(2) is located at the center of a Co hexagon, it is expected to contribute to the in-plane magnetic exchange, especially $J_d$, third-neighbor coupling across the Co hexagon.
On the other hand, Sn(1) has a large weight at the L point (non-zero momentum) and, therefore, is expected to contribute to the out-of-plane exchange coupling. 

To analyze possible Weyl points in 120$^\circ$ AFM states, we use the WannierTools package \cite{wanniertools} with the maximally-localizes Wannier functions generated
by the Wannier90 code \cite{wannier90}. 
Figure \ref{fig:weyl} shows the obtained Weyl points in $\rm R\mbox{-}3m$(top) and $\rm R-3m'$ (bottom) 120$^\circ$ AFM states. 
In both cases, the threefold rotational symmetry (${\cal C}_{3z}$) exists as expected from the magnetic symmetry.  
Yet, some deviations can be seen, especially in $\rm R\mbox{-}3m$, because the ${\cal C}_{3z}$ symmetry is slightly broken in the constrained magnetism calculations.

\begin{figure}
\begin{center}
\centering \includegraphics[width=1\linewidth]{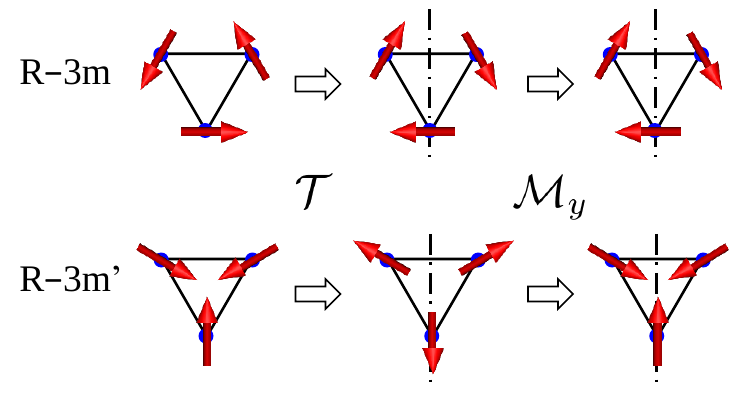}
\caption{Combined time-reversal ($\cal T$) and mirror-reflection (${\cal M}_y$) symmetry. 
While this symmetry is broken in the R$-3$m AFM state, it is preserved in the R$\mbox{-}$3m' AFM state as in the FM state. 
Thus, the R$\mbox{-}$3m' AFM state has Weyl points in momentum planes that maintain the $\cal T$-${\cal M}_y$ symmetry and 
equivalent planes related by ${\cal C}_3z$ rotation symmetry. }
\label{fig:tandmy}
\end{center}
\end{figure}

In $\rm R\mbox{-}3m$ AFM state, we found that 20 Weyl points are concentrated along the $\Gamma$-T line. 
This is because two nearly-degenerate bands exist along this line [see Fig. 4 (a) in the main text]. 
Another 6 Weyl points are located near zone boundaries. 
These locations are distinct from those of the Weyl points in $\rm R-3m'$ AFM and FM states, 
both of which have the common symmetry, the combination of time-reversal $\cal T$ and  mirror-reflection about the $y$ direction ${\cal M}_y$, 
as shown in Fig.~\ref{fig:tandmy}. 
The combination of ${\cal T}$- ${\cal M}_y$ symmetry and the inversion symmetry $\cal I$ results in 6 Weyl points in $\rm R-3m'$ AFM and FM states \cite{Liu2018,Xu2018}. 
 
During this analysis, we noticed that  the number of Weyl points in the $\rm R\mbox{-}3m$ AFM is extremely sensitive to the numerical accuracy of the constrained magnetism calculation, but the location and the sign of chiral charge when it appears are robust. 
In the specific example shown in Fig.~\ref{fig:weyl} (top), we found 20 Weyl points along the $\Gamma$-T line and 6 near zone boundaries. These Weyl points break some symmetry of the ideal  $\rm R\mbox{-}3m$ AFM ordering 
because the magnetic symmetry is slightly lowered in the numerical result of the constrained magnetism calculation. 
In particular, 6 Weyl points near zone boundaries have the ${\cal C}_{3z}$ symmetry but not the spatial inversion $\cal I$ symmetry, 
and other Weyl points expected by the $\cal I$ symmetry are missing at $k_z<0$. 
In Fig.~4 (a), we restore the $\cal I$ symmetry and added 6 Weyl points at $k_z<0$.
